\documentclass[twocolumn,showpacs,amsmath,amssymb,floatfix,aps,prb]{revtex4}
\usepackage{graphicx}

\def\sgn{{\rm sgn}}

\def\Tr{{\rm Tr}}

\begin{document}

\title{Hysteresis, transient oscillations, and nonhysteretic switching in resonantly modulated large-spin systems}

\author{C. Hicke and M.~I.~Dykman} \affiliation{Department of Physics and Astronomy,
Michigan State University, East Lansing, MI 48824}

\date{\today}

\begin{abstract}
We study the classical dynamics of resonantly modulated large-spin systems in a strong magnetic field. We show that these systems have special symmetry. It leads to characteristic nonlinear effects. They include abrupt switching between magnetization branches with varying modulating field without hysteresis and a specific pattern of switching in the presence of multistability and hysteresis. Along with steady forced vibrations the transverse spin components can display transient vibrations at a combination of the Larmor frequency and a slower frequency determined by the anisotropy energy. The analysis is based on a microscopic theory that takes into account relaxation mechanisms important for single-molecule magnets and other large-spin systems. We find how the Landau-Lifshitz model should be modified in order to describe the classical spin dynamics. The occurrence of transient oscillations depends on the interrelation between the relaxation parameters.
\end{abstract}
\maketitle

\section{Introduction}
\label{sec:Intro}

Large-spin systems have a finite but comparatively large number of quantum states. Therefore a single system can be used to study a broad range of phenomena, from purely quantum to semiclassical where the spin behaves almost like a classical top. One of the interesting features of large-spin systems is that, in a strong static magnetic field, their energy levels become almost equidistant, with level spacing close to $\hbar\omega_0$, where $\omega_0$ is the Larmor frequency. As a result, radiation at frequency $\approx \omega_0$ is resonant simultaneously for many interlevel transitions. This leads to new quantum and classical nonlinear resonant effects.

An important class of large-spin systems is single-molecule magnets (SMMs). SMMs display an extremely rich behavior and have been attracting much attention in recent years. A variety of SMMs has already been discovered and investigated, see Refs.~\onlinecite{Wernsdorfer2001,Friedman2003,Gatteschi2006} for a review, and new systems are being found \cite{Sieber2004,Evangelisti2006}. Another example of large-spin systems is provided by large nuclear spins, the interest in which has renewed in view of their possible use in quantum computing \cite{Leuenberger2003}.

In this paper we study the dynamics of large-spin systems, $S\gg 1$,  in the classical limit. We assume that the system is in a strong static magnetic field along the easy magnetization axis and in an almost resonant transverse field. For a small relaxation rate, even a weak transverse resonant field can lead to hysteresis of the response. As we show, the hysteresis is quite unusual.

It is convenient to describe the dynamics of a resonantly modulated spin in the rotating wave approximation (RWA). The corresponding analysis in the absence of relaxation has revealed a special quantum feature, an antiresonance of the response which accompanies anticrossing of quasienergy levels \cite{Hicke2007}. Quantum spin dynamics in the rotating frame bears also on the dynamics of the Lipkin-Meshkov-Glick model
\cite{Lipkin1965,Ulyanov1992,Garanin1998,Ribeiro2007}.

One may expect that the features of the coherent quantum dynamics should have counterparts in the classical spin dynamics in the presence of dissipation. As we show, this is indeed the case. The system displays an unusual behavior in a certain range of modulation parameters. This behavior is due to a special symmetry. It leads to specific features of hysteresis and to discontinuous (in the neglect of fluctuations) switching between different response branches even in the absence of hysteresis.

Classical dynamics of a large-spin system in a resonant field would be expected to have similarities with the dynamics of a modulated magnetic nanoparticle near ferromagnetic resonance. It was understood back in the 1950's \cite{Anderson1955,Suhl1957} that the response near ferromagnetic resonance becomes strongly nonlinear already for comparatively weak radiation strength due to the magnetization dependence of the effective magnetic field. Resonant response may become multivalued as a function of the modulating field amplitude \cite{Skrotskii1959a,Seagle1984}. A detailed analysis of nonlinear magnetization dynamics in uniaxial nanoparticles modulated by a strong circularly polarized periodic field was done recently \cite{Bertotti2001}. These studies as well as many other studies of magnetization dynamics in ferromagnets were based on the Landau-Lifshitz-Gilbert equation.

In contrast to magnetic nanoparticles, for large-spin systems quantum effects are substantial. A distinction which remains important in the classical limit concerns relaxation mechanisms. Spin relaxation occurs via transitions between discrete energy levels with emission, absorption, or inelastic scattering of excitations of a thermal reservoir to which the spin is coupled. Relevant relaxation mechanisms depend on the specific system but as we show, even in the classical limit relaxation is not described, generally, by the Landau-Lifshitz damping. As a result the classical spin dynamics strongly differs from the dynamics of a magnetic nanoparticle.

The microscopic analysis of relaxation is simplified in the presence of a strong static magnetic field. Here, all spin energy levels are almost equidistant. Therefore excitations of the thermal bath  emitted, for example, in transitions within different pairs of neighboring levels have almost the same energies. As a consequence, relaxation is described by a small number of constants independent of the form of the weighted with the interaction density of states of the bath, and the analysis applies for an arbitrary ratio between the level nonequidistance and their relaxational broadening \cite{DK_review84}.

We consider three relaxation mechanisms. Two of them correspond to transitions between neighboring and next neighboring spin levels, with the coupling to bosonic excitations quadratic in the spin operators. Such coupling is important, in particular, for SMMs where energy relaxation is due to phonon scattering. The theory of relaxation of SMMs was developed earlier \cite{Garanin1997a,Leuenberger2000a} and has been tested experimentally, see Refs.~\onlinecite{Bahr2007,Bal2007} and papers cited therein. We also consider coupling to a bosonic bath linear in spin operators. It leads to relaxation that in the classical limit has the form of the Landau-Lifshitz damping provided the modulation field is weak compared to the static field, as assumed in the RWA.

The typical duration of scattering events that lead to spin relaxation is often $\sim\omega_0^{-1}$. In the RWA they appear instantaneous. The operator that describes relaxation has a simple functional form, with no retardation in the ``slow" time. This is advantageous for studying the classical limit and allows us to obtain analytical results.

In the classical limit, a spin is characterized by two dynamical variables, for example, azimuthal and polar angles. In the RWA, they satisfy autonomous equations of motion, the coefficients in these equations do not depend on time. A two-variable nonlinear dissipative system can have both stationary and periodic states \cite{Guckenheimer1987}. As we show, such states indeed emerge for a resonantly modulated spin. They were predicted also for a magnetic nanoparticle with Landau-Lifshitz damping in the case of a generally nonresonant modulation \cite{Bertotti2001}.

For a spin, the occurrence of periodic states in the rotating frame critically depends on the interrelation between the relaxation parameters. In particular, these states do not emerge for a resonantly modulated spin with microscopic relaxation that reduces to the Landau-Lifshitz damping in the RWA. Moreover, quantum fluctuations lead to phase diffusion which results in decay of periodic states in the rotating frame, making the corresponding vibrations transient.

The paper is organized as follows. In Sec.~\ref{sec:Model} we introduce a model of the spin and its interaction with a thermal bath and derive the quantum kinetic equation with account taken of different relaxation mechanisms. In Sec.~\ref{sec:classical_motion} we obtain classical equations of motion and discuss the symmetry of the system. We find analytically, for weak damping, the positions of the bifurcation curves where the number of stationary states in the rotating frame changes (saddle-node bifurcations) and where periodic states are split off from stationary states (Hopf bifurcations). Sec.~\ref{sec:mu_equals_zero} describes the specific and, perhaps, most unusual feature of the system, the occurrence of Hamiltonian-like dynamics in the presence of dissipation. In Sec.~\ref{sec:linear_in_S} spin dynamics and hysteresis are described for the relation between relaxation parameters where the system does not have periodic states in the rotating frame. In Sec.~\ref{sec:quadratic_in_S} we consider the opposite case. The onset of periodic states and their stability are analyzed and the features of the hysteresis related to the occurrence of periodic states are studied. Details of the calculations are outlined in Appendix. Sec.~\ref{sec:conclusions} contains concluding remarks.

\section{The model}
\label{sec:Model}

We consider a large spin, $S\gg 1$, in a strong stationary magnetic
field along the easy axis $z$. The spin is modulated by a
transverse magnetic field with frequency $\omega_F$ close to the
Larmor frequency $\omega_0$. The Hamiltonian of the spin has the form
\begin{eqnarray}
\label{eq:hamiltonian}
 H_0 = \omega_0 S_z - \tfrac{1}{2}D S_z^2 - S_x A \cos\omega_F t \qquad (\hbar =1)
\end{eqnarray}
This Hamiltonian well describes many single-molecule magnets, including
Mn$_{12}$ crystals; $D$ characterizes the magnetic anisotropy and $A$ is the modulation amplitude. It also describes a nuclear spin, with $D$ characterizing the quadrupolar coupling energy to an electric field gradient in the crystal with an appropriate symmetry.

We assume that the Zeeman energy levels in the absence of modulation are almost equidistant. We also assume that the resonant modulation is not too strong. These conditions are met provided
\begin{eqnarray}
\label{eq:RWA}
 |\omega_0-\omega_F|, DS, A \ll \omega_0.
\end{eqnarray}
For many SMMs the inequality $DS\ll \omega_0$ is fairly demanding  and requires strong static magnetic fields; for example $D\approx 0.6$~K for Fe$_8$ (where $S=10$) \cite{Gatteschi2006}. On the other hand, for more isotropic SMMs the anisotropy is much smaller; for example, $D\approx 0.04$~K for Fe$_{17}$ where $S=35/2$, see Ref.~\onlinecite{Evangelisti2006} (our definition of $D$ differs by a factor of 2 from the definition used in the literature on SMMs). The anisotropy is usually much weaker for large-$S$ nuclei and the condition (\ref{eq:RWA}) is not restrictive.

The quantum dynamics of an isolated spin with Hamiltonian
Eq.~(\ref{eq:hamiltonian}) was considered earlier \cite{Hicke2007}.
Here we are interested in the spin dynamics in the presence of dissipation. Different dissipation mechanisms are important for different systems. For single-molecule magnets, energy dissipation is due primarily to transitions between spin energy levels accompanied by emission or absorption of phonons. The transitions between both nearest and next nearest spin levels are important \cite{Garanin1997a,Leuenberger2000a,Villain1994}. The corresponding interactions are
\begin{eqnarray}
\label{eq:interaction1}
 &H_i^{(1)}=\sum\nolimits_k V_k^{(1)}\left[\left(S_+S_z+S_z S_+\right)b_k +{\rm h.c.}\right]\\
 &H_i^{(2)}=\sum\nolimits_k V_k^{(2)}\left(S_+^2b_k +{\rm h.c.}\right),
 \qquad S_\pm=S_x\pm iS_y \nonumber,
\end{eqnarray}
where $k$ enumerates phonon modes, $b_k$ is the
annihilation operator for the k-th mode, and $V_k^{(1)}$ and $V_k^{(2)}$ are the
coupling parameters responsible for transitions between nearest and next nearest Zeeman levels. The phonon Hamiltonian is
\begin{eqnarray}
\label{eq:phonon_hamiltonian}
 H_{ph}=\sum\nolimits_k\omega_k b_k^+b_k.
\end{eqnarray}
A similar interaction Hamiltonian describes the coupling of a nuclear spin to phonons, cf. Ref.~\onlinecite{Melcher1971} and the early work \cite{Shulman1957,Taylor1959}.

Along with the interaction (\ref{eq:interaction1}) we will
consider the interaction that is linear in the spin operators,
\begin{eqnarray}
\label{eq:interaction2}
 H_i^{(3)}=\sum\nolimits_k V_k^{(3)}\left(S_+b_k +{\rm h.c.}\right).
\end{eqnarray}
Such interaction is allowed by time-reversal symmetry in the
presence of a strong static magnetic field, with the coupling
constants $V_k^{(3)}$ proportional to the off power of the field field. It can be thought of as arising from phonon-induced modulation of the spin $g$-factor.
The interaction Eq.~(\ref{eq:interaction2}) is also important for
impurity spins in magnetic crystals, in which case $b_k$ is the
annihilation operator of a magnon \cite{Izyumov1966,Krivoglaz1968a}.

\subsection{Rotating Wave Approximation}
\label{subsec:RWA}

The dynamics of a periodically modulated spin can be conveniently
described in the rotating wave approximation. To do this we make a
canonical transformation $U(t)=\exp(-i\omega_FS_zt)$. The
transformed Hamiltonian $H_0$ then becomes $\tilde
H_0=U^{\dagger}H_0U-iU^{\dagger}\dot U$,
\begin{eqnarray}
\label{eq:hamiltonian_RWA}
&\tilde H_0=-\delta \omega S_z - \frac{1}{2}D S_z^2-\frac{1}{2}AS_x,\nonumber\\
&\delta \omega = \omega_F - \omega_0.
\end{eqnarray}
Here we disregarded fast-oscillating terms $\propto A \exp(\pm 2i\omega_Ft)$.

We note that the Hamiltonian (\ref{eq:hamiltonian_RWA}) has the form
of a free energy of a magnetic moment in an easy axis ferromagnet,
with ${\bf S}$ playing the role of the magnetization and
$\delta\omega$ and $A$ giving the components of the effective
magnetic field (in energy units) along the $z$ and $x$ axes, respectively.

It is convenient to change to dimensionless variables and rewrite the Hamiltonian as $\tilde
H_0=S^2D(\hat g+\mu^2/2)$, with
\begin{eqnarray}
\label{eq:hamiltonian_g}
 &&\hat g = -\frac{1}{2}(s_z +\mu)^2-fs_x,\nonumber\\
 &&{\bf s} = {\bf S}/S,\qquad \mu=\delta\omega/SD,\qquad f=A/2SD.
\end{eqnarray}
The Hamiltonian $\hat g$ describes the dynamics of an isolated spin in ``slow" dimensionless time $\tau = SDt$. It gives dimensionless quasienergies of a periodically modulated spin in the RWA.  From Eq.~(\ref{eq:hamiltonian_g}), the spin dynamics is determined by the two dimensionless parameters, $\mu$ and $f$, which depend on the interrelation between the frequency detuning of the modulating field $\delta\omega$, the anisotropy parameter $DS$, and the modulation amplitude $A$. The spin variables $\hat{\bf s}$ are advantageous for describing large spins, since the commutators of their components are $\propto S^{-1}$, which simplifies a transition to the classical limit for $S\gg 1$.

\subsection{Quantum kinetic equation}
\label{subsec:QKE}

We will assume that the interaction with phonons (magnons) is weak. Then under standard conditions the equation of motion for the spin density matrix $\rho$ is Markovian in slow time $\tau$, i.e., on a time scale that largely exceeds $\omega_F^{-1}$ and the typical correlation time of  phonons (magnons). We will switch to the interaction representation with respect to the Hamiltonian $\omega_FS_z + H_{\rm ph}$. Then to leading order in the spin to bath coupling the quantum kinetic equation can be written as
\begin{eqnarray}
\label{eq:QKE}
S^{-1} \dot \rho = i[\rho,g] -\hat\Gamma^{(1)}\rho -\hat\Gamma^{(2)}\rho -\hat\Gamma^{(3)}\rho,
\end{eqnarray}
where $\dot A \equiv dA/d\tau$.

The operators $\hat\Gamma^{(j)}$ describe relaxation due to the interactions $H_i^{(j)}$, with $j=1,2,3$. They can be written schematically as
\begin{eqnarray}
\label{eq:QKE_Relaxation}
\hat\Gamma\rho&=&\Gamma[
\left(\bar{n}+1\right)\left(L^+L\rho-2L\rho L^++\rho L^+L \right)\nonumber\\
&&+\bar{n}\left(L L^+\rho-2L^+\rho L+\rho L L^+ \right)]
\end{eqnarray}
Here we have taken into account that all transitions between spin
states with emission or absorption of phonons (magnons) involve
almost the same energy transfer $\Delta E$, with $\Delta E\approx\omega_F$ for terms $\propto \Gamma^{(1)}, \Gamma^{(3)}$ and $\Delta E\approx 2\omega_F$ for the term $\propto\Gamma^{(2)}$. In this sense,
equation for spin relaxation (\ref{eq:QKE_Relaxation}) resembles the
quantum kinetic equation for a weakly nonlinear oscillator coupled
to a bosonic bath \cite{DK_review84}. Respectively, $\bar{n}$ is
the Planck number of the emitted/absorbed phonons,
$\bar{n}=\left[\exp(\Delta E/kT-1)\right]^{-1}$. Because of the same
transferred energy, different transitions are characterized by the
same rate constants, which for the interactions $H_i^{(1)-(3)}$ have the following form, in dimensionless time:
\begin{eqnarray}
\label{eq:rate_constants}
 &&\Gamma^{(1)}=\pi D^{-1}
S^2\sum\nolimits_k\left|V_k^{(1)}\right|^2
\delta\left(\omega_F-\omega_k\right),\nonumber\\
 &&\Gamma^{(2)}=\pi D^{-1}
S^2\sum\nolimits_k\left|V_k^{(2)}\right|^2
\delta\left(2\omega_F-\omega_k\right),\nonumber\\
&&\Gamma^{(3)}=\pi D^{-1}
\sum\nolimits_k\left|V_k^{(3)}\right|^2\delta\left(\omega_F-\omega_k\right).
\end{eqnarray}
The operators $L$ for the interactions $H_i^{(1)-(3)}$
are
\begin{eqnarray}
\label{eq:L_operators}
 L^{(1)}=s_-s_z+s_z s_-,\quad L^{(2)}=s_-^2, \quad
L^{(3)}=s_-,
\end{eqnarray}
where $s_{\pm}=S_{\pm}/S$.

It is important to note that, along with dissipation, coupling to phonons (magnons) leads to a polaronic effect of renormalization of the spin energy. A standard analysis shows that renormalization due to $H_i^{(3)}$, to second order in $H_i^{(3)}$ comes to a change of the anisotropy parameter $D$ and the Larmor frequency. A similar change comes from nonresonant terms $\propto S_+b_k^{\dagger} + H.c.$. In contrast, renormalization from $H_i^{(1),(2)}$, along with terms $\propto S_z,S_z^2$, leads to terms of higher order in $S_z$ in the spin Hamiltonian, in particular to terms $\propto S_z^4$. The condition that they are small compared to the anisotropy energy $D S_z^2$ imposes a constraint on the strength of the coupling $H_i^{(1),(2)}$. Respectively, we will assume that the dimensionless decay rates $\Gamma^{(1),(2)}$ are small, $\Gamma^{(1),(2)}\ll 1$. It is not necessary to impose a similar condition on the dimensionless rate $\Gamma^{(3)}$. Still we will be interested primarily in the spin dynamics in the underdamped regime, where $\Gamma^{(1)-(3)}$ are small.

\section{Classical motion of the modulated spin}
\label{sec:classical_motion}

The analysis of spin dynamics is significantly simplified in the classical, or mean-field limit. Classical equations of motion for the spin components can be obtained by multiplying Eq.~(\ref{eq:QKE}) by $s_i$ $(i=x,y,z)$, taking the trace, and decoupling $\Tr\left(s_{i_1} s_{i_2}\rho\right)\to s_{i_1} s_{i_2}$. The decoupling should be done after the appropriate commutators are evaluated; for example, $\Tr\left(\left[s_z,\hat g\right]\rho\right)\to -ifs_y$. From Eqs.~(\ref{eq:hamiltonian_g}), (\ref{eq:QKE}), (\ref{eq:L_operators}) we obtain
\begin{eqnarray}
\label{eq:classical}
&&\dot {\bf s} = -{\bf s}\times \partial_{\bf s} g + \left(\dot{\bf s}\right)_d, \quad \left(\dot{\bf s}\right)_d= \Gamma_d(s_z){\bf s}\times \left({\bf s}\times \hat{\bf z}\right),\nonumber\\
&&\Gamma_d(s_z)=2\left(4\Gamma^{(1)}s_z^2 +2\Gamma^{(2)}(1-s_z^2) +\Gamma^{(3)}\right),
\end{eqnarray}
where $\hat{\bf z}$ is a unit vector along the $z$-axis, which is the direction of the strong dc magnetic field.

We have assumed in Eq.~(\ref{eq:classical}) that $S\gg \bar n$. Note that, in dimensional units, $S= |{\bf L}|/\hbar$, where ${\bf L}$ is the angular momentum, whereas in the classical temperature limit $\bar n =kT/\hbar\omega_F$ or $kT/2\hbar\omega_F$ depending on the scattering mechanism. Therefore the condition $S\gg \bar n$ imposes a limitation on temperature.

Equation~(\ref{eq:classical}) is reminiscent of the Landau-Lifshitz equation for magnetization of a ferromagnet. However, in contrast to the Landau-Lifshitz equation a retardation-free equation of motion for a classical spin could be obtained only in the rotating frame, that is, in slow time $\tau$. The term with $\partial_{\bf s} g$ describes precession of a spin with energy (quasienergy, in the present case) $g$. The term $\left(\dot{\bf s}\right)_d$ describes the effective friction force. It is determined by the instantaneous spin orientation, but its form is different from that of the friction force in the Landau-Lifshitz equation.

We emphasize that Eq.~(\ref{eq:classical}) is not phenomenological, it is derived for the microscopic model of coupling to the bath (\ref{eq:interaction1}), (\ref{eq:interaction2}). We now consider what would happen if we start from the Landau-Lifshitz equation and switch to the rotating frame using the RWA in the assumption that the resonant driving is comparatively weak, $A\ll \omega_0$ [cf. Eq.~(\ref{eq:RWA})]. In this case one should keep in the expression for the friction force only the leading term in the effective magnetic field, i.e., assume that ${\bf H}\parallel \hat{\bf z}$. The result would be Eq.~(\ref{eq:classical}) with a dissipative term of the same form as the term $\propto \Gamma^{(3)}$ but without dissipative terms that have the structure of the terms $\propto\Gamma^{(1)}, \Gamma^{(2)}$. However, these latter terms play a major role for SMMs \cite{Garanin1997a,Leuenberger2000a,Bahr2007,Bal2007} and for phonon scattering by nuclear spins.

As mentioned in the Introduction, the dynamics of a single-domain magnetic nanoparticle in a circularly polarized field was studied using the Landau-Lifshitz-Gilbert equation in a series of papers \cite{Bertotti2001}. It is clear from the above comment that the results of this analysis do not generally describe resonant behavior of SMMs. Moreover, periodic states in the rotating frame predicted in Ref.~\onlinecite{Bertotti2001} do not arise in resonantly excited spin systems with relaxation $\propto\Gamma^{(3)}$, as shown below.

\subsection{Stationary states in the rotating frame for weak damping}
\label{subsec:stationary_general}

A classical spin is characterized by its azimuthal and polar angles, $\phi$ and $\theta$, with $s_z=\cos\theta, s_x=\sin\theta\cos\phi, s_y=\sin\theta\sin\phi$. In canonically conjugate variables $\phi,s_z$ equations of motion (\ref{eq:classical}) take the form
\begin{eqnarray}
\label{eq:classic_sz_phi}
 \dot \phi&=& \partial_{s_z} g,\\
 \dot s_z&=&-\partial_{\phi} g -\Gamma_d(s_z)(1-s_z^2),\nonumber
\end{eqnarray}
where $g$ as a function of $s_z,\phi$ has the form $g=-(s_z+\mu)^2/2-f(1-s_z^2)^{1/2}\cos\phi$, cf. Eq.~(\ref{eq:hamiltonian_g}). We note that the dissipation term is present only in the equation for $\dot s_z$.

\begin{figure}[h]
\includegraphics[width=3.2in]{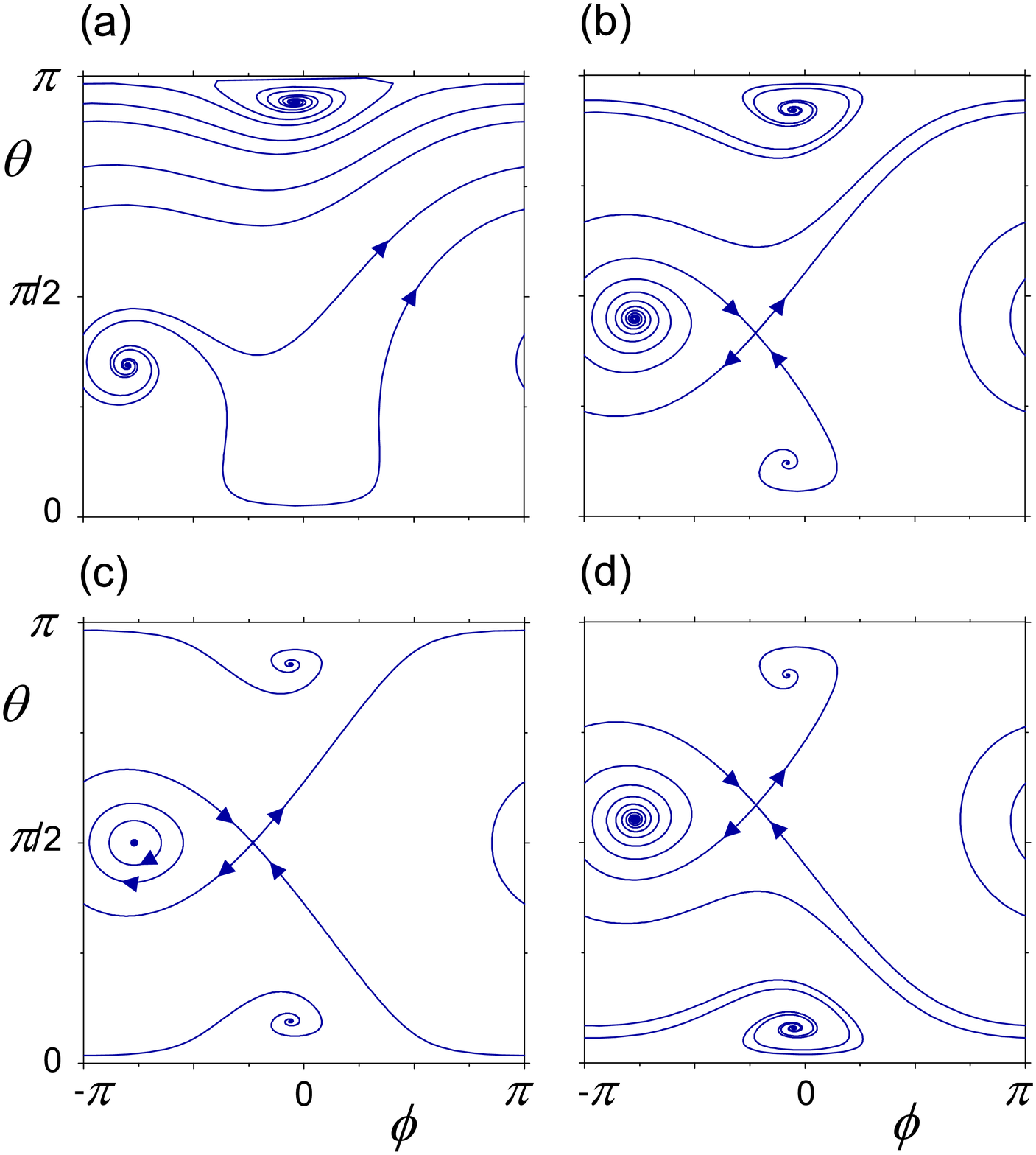}
\caption{(Color online). Phase portraits of the spin on $(\theta,\phi)$-plane ($s_z=\cos\theta$). The data refer to $\Gamma^{(1)}=\Gamma^{(2)}=0$,  $\Gamma^{(3)}=0.1$, and $f=0.3$. In panels (a)-(d) $\mu=-0.6,-0.2, 0, 0.2$, respectively.}
\label{fig:small_gamma_2_traj}
\end{figure}

In the absence of relaxation, precession of a spin with given $g$ corresponds to moving along orbits on the $(\phi,s_z)$-plane. The orbits can be either closed or open; in the latter case $\phi$ varies by $2\pi$ over a period, cf. Fig.~\ref{fig:small_gamma_2_traj}. There are also stationary states where the spin orientation does not vary in time. Generally, relaxation breaks this structure. If it is weak it makes some of the stationary states asymptotically stable or unstable and can also transform some of the orbits into stable or unstable limit cycles, which correspond to periodic oscillations of $s_z$ and $\phi$ in the rotating frame. The frequency of these oscillations is determined by the system nonlinearity and is not immediately related to a combination of the modulation frequency and the Larmor frequency, for example.

Stationary states of Eq.~(\ref{eq:classic_sz_phi}), which is written in the rotating frame, correspond to the states of forced vibrations of the spin components $s_x,s_y$ at frequency $\omega_F$ in the laboratory frame. Periodic states in the rotating frame correspond, in the laboratory frame, to periodic vibrations of $s_z$ and to vibrations of $s_x,s_y$ at combination frequencies equal to $\omega_F$ with added and subtracted multiples of the oscillation frequency in the rotating frame (which is small compared to $\omega_F$). In what follows we keep this correspondence in mind, but the discussion refers entirely to the rotating frame.

The analysis of stability of stationary states is based on linearizing the equations of motion near these states and looking at the corresponding eigenvalues $\lambda_1,\lambda_2$ \cite{Guckenheimer1987}. In the absence of damping the stationary states are either hyperbolic points (saddles) with real $\lambda_{1,2}$ or elliptic points (centers) with imaginary $\lambda_{1,2}$. From Eq.~(\ref{eq:classic_sz_phi}), a fixed point is hyperbolic if $\lambda_1\lambda_2={\cal D} < 0$, where
\begin{equation}
\label{eq:determinant}
 {\cal D} = \partial_{\phi}^2g\,\partial_{s_z}^2g - \left(\partial_{\phi}\partial_{s_z}g\right)^2
\end{equation}
(the derivatives are calculated at the stationary state). On the other hand, if ${\cal D} > 0$ the stationary state corresponds to an elliptic point, orbits $g=$~const are circling around it.

For weak damping, hyperbolic points remain hyperbolic. On the other hand, a center becomes asymptotically stable (an attractor) or unstable (a repeller) for ${\cal T} < 0$ or ${\cal T} >0$, respectively. Here ${\cal T}=-\partial [\Gamma_d(s_z)(1-s_z^2)]/\partial s_z$, or in explicit form
\begin{equation}
\label{eq:trace}
 {\cal T}=-4s_z\left[4\Gamma^{(1)}(1-2s_z^2) -4\Gamma^{(2)}(1-s_z^2) - \Gamma^{(3)}\right],
 \end{equation}
where $s_z$ is taken for the appropriate center;  $\lambda_1+\lambda_2={\cal T}$. The sign of ${\cal T}$ determines stability of a stationary state also where dissipation is not small.

The quasienergy $g$ has symmetry properties that the change $f\to -f$ can be accounted for by replacing $\phi\to\phi +\pi, s_z\to s_z$. This replacement preserves the form of equations of motion (\ref{eq:classic_sz_phi}) also in the presence of damping. Therefore in what follows we will concentrate on the range $f\geq 0$. On the other hand, the change $\mu\to -\mu$ would not change $g$ if we simultaneously replace $\phi\to \phi, s_z\to -s_z$. In equations of motion one should additionally change $\tau \to -\tau$. Therefore, if for $\mu =\mu^{(0)}< 0$ the system has an attractor located at a given $(\phi^{(0)}, s_z^{(0)})$, then for $\mu=-\mu_0$ it has a repeller located at $\phi^{(0)}, -s_z^{(0)}$. This behavior is illustrated in Fig.~\ref{fig:small_gamma_2_traj}, where panels (b) and (d) refer to opposite values of $\mu$.

\subsection{Saddle-node bifurcations}
\label{subsec:saddle-node}

\begin{figure}[h]
\includegraphics[width=3.2in]{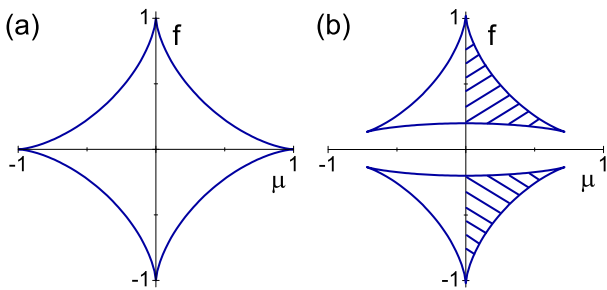}
\caption{(Color online). Saddle-node bifurcation lines. Panel (a):  zero-damping limit, the lines have the form of the Stoner-Wolfarth astroid in the variables of reduced amplitude $f$ and frequency detuning $\mu$ of the resonant field. Panel (b): nonzero damping, $\Gamma^{(3)}=0.1$, $\Gamma^{(1)}=\Gamma^{(2)}=0$. In the dashed region the spin has two coexisting stable equilibria in the rotating frame. }
\label{fig:astroid}
\end{figure}

The function $g({\bf s})$ has a form of the free energy of a magnetic moment of an easy axis ferromagnet, as mentioned earlier, with $\mu$ and $f$ corresponding to the components of the magnetic field along and transverse to the easy axis, respectively. It is well known that $g$ may have either two or four extreme points where $\partial g/\partial s_z=\partial g/\partial \phi =0$. The region where there are four extrema lies inside the Stoner-Wohlfarth astroid \cite{Stoner1948} $|f|^{2/3}+|\mu|^{2/3}=1$ on the plane of the dimensionless parameters $\mu$ and $f$, see Fig.~\ref{fig:astroid}(a). The extrema of $g$ outside the astroid are a minimum and a maximum, whereas inside the astroid $g$ additionally has a saddle and another minimum or maximum. All of them lie at $\phi=0$ or $\phi=\pi$.

In the presence of weak damping, the minima and maxima of $g$ become stable or unstable stationary states. We note that there are no reasons to expect that the stable states lie at the minima of $g$, because $g$ is not an energy but a quasienergy of the spin. The number of stable/unstable stationary states changes on the saddle-node bifurcation curve on the $(f,\mu)$-plane. The condition that two stationary states merge \cite{Guckenheimer1987} has the form
\begin{eqnarray}
\label{eq:saddle_node_gamma_2}
{\cal D} + {\cal T}\partial_{\phi}\partial_{s_z}g = 0.
\end{eqnarray}
For weak damping a part of the curve given by this equation is close to the astroid. On the astroid $s_z=-\sgn (\mu)|\mu|^{1/3}$. Then from Eq.~(\ref{eq:trace}) for the merging saddle and node
\begin{eqnarray}
\label{eq:trace_on_astroid}
&&{\cal T}=-4\sgn(\mu)\sqrt{1-|f|^{2/3}}\nonumber\\
&&\quad\times \left(4\Gamma^{(1)}(1-2|f|^{2/3})+4\Gamma^{(2)}
 |f|^{2/3}+\Gamma^{(3)}\right).
\end{eqnarray}
If damping $\propto \Gamma^{(1)}$ is weak, the node is stable for $\mu > 0$ and unstable for $\mu <0$. On the other hand, if $\Gamma^{(1)}$ is large, the stability depends on the value of $f$.

The most significant difference between the saddle-node bifurcation curve and the Stoner-Wohlfarth astroid is that the bifurcation curve consists of two curvilinear triangles, that is, the astroid is ``split", see Fig.~\ref{fig:astroid}(b) and Fig.~\ref{fig:astroid_gamma1} below. This is also the case for a modulated magnetic nanoparticle \cite{Bertotti2001}. The triangles are obtained from Eqs.~(\ref{eq:classic_sz_phi}) and (\ref{eq:saddle_node_gamma_2}). After some algebra we find that the ``bases" of the bifurcation triangles are given by expression
\begin{equation}
\label{eq:astroid_bases}
f_B\approx \pm \Gamma_d(\mu)(1-\mu^2)^{1/2},
\end{equation}
to leading order in $\Gamma_d$. This expression applies not too close to the vertices of the triangles. We note, however, that Eq.~(\ref{eq:astroid_bases}) gives the exact bifurcational value of $f_B$ for $\mu=0$ and arbitrary $\Gamma_d(0)$.

The shape of the gap between the upper and lower curvilinear bifurcation triangles depends on the damping mechanism. In particular, the damping $\propto \Gamma^{(1)}$ does not contribute to the gap for small $|\mu|$ (cf. Fig.~\ref{fig:astroid_gamma1}), whereas the damping $\propto \Gamma^{(2)}$ does not contribute to the gap at small $1-|\mu|$. The damping-induced change of the sides of the triangles compared to the astroid is quadratic in $\Gamma_d$, far from the small-$f$ range.

The positions of the small-$f$ vertices of the bifurcation triangles $f_C, \mu_C$ for small damping can be found from Eqs.~(\ref{eq:classic_sz_phi}) and the condition that Eq.~(\ref{eq:saddle_node_gamma_2}) has a degenerate root, which gives
\begin{eqnarray}
\label{eq:critical_f_mu}
&&\mu_C\approx \pm\left[1-\sqrt{3}\left(-\Gamma_d^2 + {\cal T}\Gamma_d\right)^{1/2}\right],\nonumber\\
&&f_C \approx \pm (64/27)^{1/4}\,\Gamma_d^{3/4}\left(\Gamma_d + (1/2){\cal T}\right)^{1/2}\left(-\Gamma_d +{\cal T}\right)^{1/4},\nonumber
\end{eqnarray}
where $\Gamma_d$ and ${\cal T}$ are calculated for $s_z=1$.

\subsection{Periodic states and Hopf bifurcations}
\label{subsec:hopf_bif_general}

An important property of the modulated classical spin is the possibility to have periodic states in the rotating frame. Such states result from  Hopf bifurcations in which a stationary state transforms into a limit cycle \cite{Guckenheimer1987}. A Hopf bifurcation  occurs if
\[{\cal T}=0, \qquad {\cal D} > 0\]
in the stationary state. Besides the special case $s_z=0$ discussed in Sec.~\ref{sec:mu_equals_zero}, the corresponding stationary state is at $s_z= s_{zH}$, where
\begin{eqnarray}
\label{eq:Hopf_general}
 && s_{zH}=\alpha \frac{1}{2}\left(\frac{4\Gamma^{(1)}-4\Gamma^{(2)}-\Gamma^{(3)}} {2\Gamma^{(1)}-\Gamma^{(2)}}\right)^{1/2}, \\
&&\alpha=\pm 1,\qquad \Gamma^{(1)} \geq \Gamma^{(2)}+\frac{1}{4}\Gamma^{(3)} \nonumber
\end{eqnarray}
(the inequality on the damping parameters follows from the condition $(s_z^2)_H \leq 1$)

The field $f_H$ on the Hopf bifurcation lines as a function of the reduced detuning $\mu$ is given by a particularly simple expression for weak damping. In this case, from second equation (\ref{eq:classic_sz_phi}) the phase $\phi_H$ for the bifurcating stationary state is close to either $0$ or $\pi$ with the additional constraint $\partial^2_{s_z}g\,\partial^2_{\phi}g > 0$. Then from first equation (\ref{eq:classic_sz_phi}) and Eq.~(\ref{eq:Hopf_general}) we find that Hopf bifurcation curves are straight lines, in the limit of vanishingly small damping,
\begin{eqnarray}
\label{eq:Hopf_through_s_H}
 &&f_H=\pm\left[1- s_{zH}^2\right]^{1/2}\left[1+ \mu s_{zH}^{-1}\right], \\
&&|f_H| \geq\left[1- s_{zH}^2\right]^{3/2} \quad {\rm or} \quad
 |\mu| \geq \left| s_{zH}\right|.
\end{eqnarray}
The structure of these lines is seen in Fig.~\ref{fig:astroid_gamma1} below. They end on the saddle-node bifurcation curves and are tangent to these curves at the end points. A detailed analysis is presented in Sec.~\ref{sec:quadratic_in_S}.

\section{Hamiltonian-like motion at exact resonance}
\label{sec:mu_equals_zero}

The spin dynamics (\ref{eq:classical}) displays an unusual and unexpected behavior where the modulation frequency $\omega_F$ coincides with the Larmor frequency $\omega_0$, in which case $\mu=0$. This is a consequence of the symmetry of the quasienergy and the dissipation operator. In a certain range of dynamical variables $\phi, s_z$, the spin behaves as if there were no dissipation, even though dissipation is present. This behavior is seen in the pattern of phase trajectories of the spin. An example of the pattern is shown in Fig.~\ref{fig:small_gamma_2_traj}(c) for the case $\Gamma^{(2)} = \Gamma^{(3)} = 0$, but the behavior is not limited to this case. As seen from Fig.~\ref{fig:small_gamma_2_traj}(c), phase trajectories form closed loops, typical for Hamiltonian systems.

For $|f|$ lying inside the bifurcation triangles, the Hamiltonian-like dynamics occurs only in a part of the phase plane. This region of $f$ corresponds to $\Gamma_d(0) < |f| < [1+\Gamma_d^{2}(0)]^{1/2}$   [the upper bound on $|f|$ for $\mu=0$ can be easily obtained from Eqs.~(\ref{eq:classic_sz_phi}), (\ref{eq:saddle_node_gamma_2})]. Here, the spin has four stationary states. For small $|\mu|$ two of them have small $|s_z|$, $s_z\approx -\mu/(1-f\cos \phi)$ where $\sin\phi\approx -\Gamma_d(0)/f$. One of these states is a saddle point [$\phi \approx -\arcsin[\Gamma_d(0)/f]$] and the other is a focus [$\phi \approx \pi+\arcsin[\Gamma_d(0)/f]$].

For $\mu=0$ there occurs a global bifurcation, a homoclinic saddle-saddle bifurcation (saddle loop \cite{Guckenheimer1987}) where the separatrix coming out from the saddle goes back into it, forming a homoclinic orbit. Simultaneously, the focus inside the loop becomes a center,  ${\cal T}=0$ for $s_z=0$. All trajectories inside the homoclinic orbit are closed loops. In contrast to the case of the vicinity of the double-zero eigenvalue bifurcation \cite{Guckenheimer1987}, the pattern persists throughout a broad region of $f$.

We show how a Hamiltonian-like region in phase space emerges first for weak damping. For $\mu=0$ the quasienergy $g$ corresponds to the Hamiltonian of a spin with anisotropy energy $\propto S_z^2$, which is in a transverse field $\propto f$. Such spin in quantum mechanics has special symmetry, it can be mapped onto a particle in a symmetric potential \cite{Ulyanov1992,Garanin1998}. A part of the classical $g=$~const orbits are closed loops on the $(\phi, s_z)$-plane. They surround the center $(s_z=0,\phi=\pi)$. The orbits are symmetric with respect to the replacement
\begin{equation}
\label{eq:symmetry}
s_z\to -s_z,\;\phi\to \phi, \;
\end{equation}
which leads to $\dot\phi \to -\dot\phi, \; \dot s_z\to \dot s_z$.

Weak damping would normally cause drift of quasienergy. The drift velocity averaged over the period $\tau_p(g) $ of motion along the orbit is
\begin{equation}
\label{eq:energy_change}
 \langle \dot g\rangle = -\tau_p^{-1}\int\nolimits_0^{\tau_p} d\tau \partial_{s_z}g\,\Gamma_d(s_z)(1-s_z^2).
 \end{equation}
From the symmetry (\ref{eq:symmetry}) and the relation $\Gamma_d(s_z)=\Gamma_d(-s_z)$, we have $\langle \dot g\rangle = 0$ on a closed orbit for $\mu=0$. Therefore a closed orbit remains closed to first order in $\Gamma_d$. Of course, for open orbits, where $\phi$ is incremented by $2\pi$ over a period, $\langle \dot g\rangle \neq 0$. These orbits become spirals in the presence of damping.

Spirals and closed orbits should be separated by a separatrix, which must be a closed orbit itself. Since the separatrix must start and end at the saddle point, we understand that at $\mu=0$ for small $\Gamma_d$ there occurs a saddle-saddle homoclinic bifurcation.

The topology discussed above persists as $\Gamma_d$ increases. The symmetry (\ref{eq:symmetry}) is not broken by $\Gamma_d$. Indeed, from equations of motion (\ref{eq:classic_sz_phi}), any orbit that crosses $s_z=0$ twice per period for $\mu=0$ has the property (\ref{eq:symmetry}) and therefore is closed. The closed orbits surround the center $s_z=0,\phi=\pi-\arcsin(\Gamma_d(0)/f)$ and fill out the whole interior of the separatrix loop.

The Hamiltonian-like behavior is displayed also for $\mu=0$ and $f$ lying outside the bifurcation triangles. Here, the system has two stationary states, both with $s_z=0$ but with different $\phi$. From Eq.~(\ref{eq:trace}), for both of them ${\cal T}$ changes sign as $\mu$ goes through zero. Because there is no saddle point, for small $|\mu|$ there is no separatrix, trajectories spiral toward or away from stationary states and possibly limit cycles. It follows from the arguments above that for $\mu=0$ all trajectories become closed orbits. This is confirmed by numerical calculations for different relaxation mechanisms.

It is convenient to analyze the overall dynamics of the spin system for $\mu\neq 0$ separately for the cases where the system does or does not have stable periodic states in the rotating frame. In turn, this is determined by the interrelation between the damping parameters, cf. Eq.~(\ref{eq:Hopf_general}). Such analysis is carried out in Secs.~\ref{sec:linear_in_S} and \ref{sec:quadratic_in_S}.

\section{Spin dynamics in the absence of limit cycles}
\label{sec:linear_in_S}

We start with the case where the system does not have limit cycles. It corresponds to the situation where the damping parameter $\Gamma^{(1)}$ is comparatively small and the interrelation between the damping parameters (\ref{eq:Hopf_general}) does not hold. To simplify the analysis we set $\Gamma^{(1)}=\Gamma^{(2)}=0$, i.e., we assume that the coupling to the bath is linear in the spin operators and is described by the interaction Hamiltonian $H_i^{(3)}$. The qualitative results of this Section apply also for nonzero $\Gamma^{(1)},\Gamma^{(2)}$ as long as $\Gamma^{(3)}+4\Gamma^{(2)}>4\Gamma^{(1)}$.  The bifurcation diagram for this case is shown in Fig.~\ref{fig:astroid}.

From the form of the function ${\cal T}$, Eq.~(\ref{eq:trace}), it follows that the damping $\propto \Gamma^{(3)}$ transforms centers of conservative motion with $s_z>0$ into unstable foci (repellers), whereas centers with $s_z < 0$ are transformed into stable foci (attractors). Therefore for $\mu<0$ the spin has one stable state. It also has one stable state in the unshaded region of the half-plane $\mu>0$ (outside the bifurcation triangles in Fig.~\ref{fig:astroid}). Inside the shaded regions within the triangles the spin has two coexisting stable states.

Examples of the phase portrait are shown in Fig.~\ref{fig:small_gamma_2_traj}. As expected, for weak damping the system has a stable and an unstable focus outside the bifurcation triangles, Fig.~\ref{fig:small_gamma_2_traj}(a). In the shaded region inside the triangle it has two stable foci, an unstable
focus, and a saddle point, Fig.~\ref{fig:small_gamma_2_traj}(d). In the unshaded region inside the triangle there is one stable and two unstable foci, Fig.~\ref{fig:small_gamma_2_traj}(b) (the values of $\mu$ in panels (b) and (d) differ just by sign).

\subsection{Hysteresis of spin response in the absence of limit cycles}
\label{subsec: hysteresis_gamma2}

\begin{figure}[h]
\includegraphics[width=2.4in]{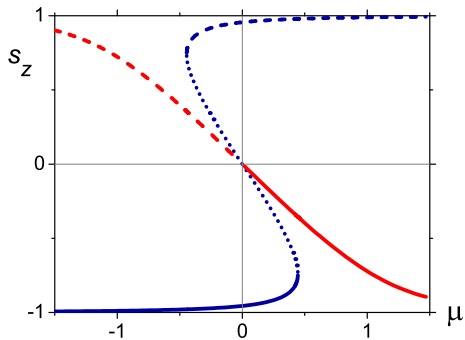}
\caption{(Color online). Hysteresis of spin response in the absence of periodic states in the rotating frame. The data refer to $\Gamma^{(1)} =\Gamma^{(2)}=0$, $\Gamma^{(3)}=0.1$, and $f=0.3$. The solid and dashed lines show, respectively, stable and unstable stationary states, the dotted line shows the saddle point.}
\label{fig:hysteresis_gamma_2}
\end{figure}

The presence of two coexisting stable states leads to hysteresis of the spin response to the external field. Such hysteresis with varying dimensionless parameter $\mu$, which is proportional to the detuning of the field frequency, is shown in Fig.~\ref{fig:hysteresis_gamma_2}. For large negative $\mu$ the system has one stable state with negative $s_z$, cf. Fig.~\ref{fig:small_gamma_2_traj}(a). As $\mu$ increases the system stays on the corresponding branch (the lowest solid line in Fig.~\ref{fig:hysteresis_gamma_2}) until the stable state merges with the saddle point (the saddle-node bifurcation). This happens for $\mu>0$ as $\mu$ reaches the bifurcation triangle. As $\mu$ further increases the system switches to the branch with larger $s_z$ and then moves along this branch. If $\mu$ decreases starting with large values where the system has only one stable state, the switching to the second branch occurs for $\mu=0$.

The hysteresis pattern in Fig.~\ref{fig:hysteresis_gamma_2} differs from the standard $S$-shape characteristic. This is the case for any  $f$ lying between the minimum and maximum of the bifurcation triangle for $\mu=0$, i.e., for $2\Gamma^{(3)}< |f| < \left(1+4\Gamma^{(3)\,2}\right)^{1/2}$. It is a consequence of the fact that the bifurcation at $\mu=0$ is not a saddle-node bifurcation, whereas a most frequently considered $S$-shape hysteresis curve arises if both bifurcations are of the saddle-node type. In our case, for $\mu=0$ the branch which is stable in the range of large positive $\mu$ (the upper stable branch in Fig.~\ref{fig:hysteresis_gamma_2}) becomes unstable as a result of the motion becoming Hamiltonian-like. The value of $s_z$ on this branch for $\mu=0$ is $s_z=0$, it coincides with the value of $s_z$ at the saddle (but the values of $s_x$ are different). Therefore when $s_z$ is plotted as a function of $\mu$ the branch, which is stable for large positive $\mu$ crosses with the branch that corresponds to the saddle point. For negative $\mu$ the branch, which is stable for large positive $\mu$, becomes unstable, cf. Fig.~\ref{fig:small_gamma_2_traj}. As $\mu$ decreases and reaches the bifurcation triangle for $\mu<0$, the saddle  merges with an unstable equilibrium as seen in Fig.~\ref{fig:hysteresis_gamma_2}.

The spin components display hysteresis also if the shaded area of the bifurcation triangle in Fig.~\ref{fig:astroid}(b) is crossed in a different way, for example, by varying $f$. 
If the crossing occurs so that the line $\mu=0$ is not crossed, the hysteresis curves have a standard $S$ shape. We note that hysteresis of $s_x, s_y$ corresponds to hysteresis of amplitude and phase of forced vibrations of the spin.

\subsection{Interbranch switching without hysteresis}
\label{subsec:switching_no_hysteresis}

\begin{figure}[h]
\includegraphics[width=2.4in]{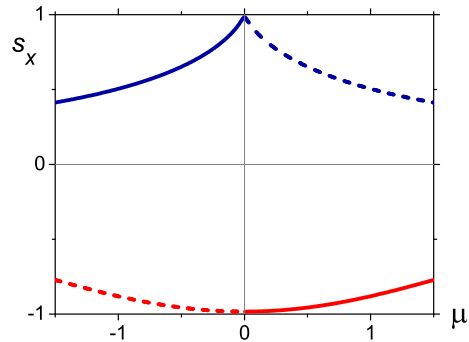}
\caption{(Color online). Frequency dependence of the transverse spin component for field amplitudes $f$ where the system has one stable state. The solid and dashed lines show the stable and unstable values of $s_x$ in the rotating frame. The data refer to $\Gamma^{(1)} =\Gamma^{(2)}=0$, $\Gamma^{(3)}=0.1$, and $f=1.1$. As the scaled frequency detuning $\mu$ goes through $\mu=0$ the value of $s_x$ changes to almost opposite in sign.}
\label{fig:s_x_switch}
\end{figure}

The occurrence of Hamiltonian dynamics for $\mu=0$ leads to an interesting and unusual behavior of the system even outside the bifurcation triangles, i.e. in the region where the system has only one stable state. In the small damping limit and for $|f|>1$ and $|\mu|\ll 1$ the stationary states are at $\phi = 0$ and $\phi=\pi$, with $s_z=\mu/(f\cos\phi-1)$. The stable state is the one with $s_z < 0$, whereas the one with $s_z>0$ is unstable. As $\mu$ goes through zero the states with $\phi=0$ and $\phi=\pi$ interchange stability. This means that $s_x\approx \cos\phi$ jumps between $-1$ and $1$. Such switching is seen in Fig.~\ref{fig:s_x_switch}.

\section{Spin dynamics in the presence of limit cycles}
\label{sec:quadratic_in_S}

The classical dynamics of the spin changes significantly if the spin has stable periodic states in the rotating frame. This occurs where condition (\ref{eq:Hopf_general}) on the damping parameters is met. The features of the dynamics can be understood by setting $\Gamma^{(2)}=\Gamma^{(3)}=0$, $\Gamma^{(1)}>0$, i.e., by assuming that damping is due primarily to coupling to a bath $H_i^{(1)}$, which is quadratic in spin components, with elementary scattering processes corresponding to transitions between neighboring Zeeman levels. This model is of substantial interest for single-molecule magnets \cite{Garanin1997a,Bal2007}.
\begin{figure}[h]
\includegraphics[width=3.2in]{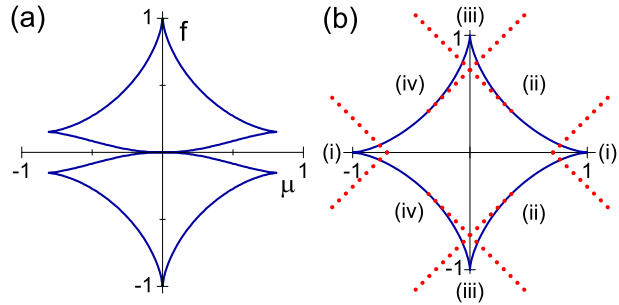}
\caption{(Color online). (a) Saddle-node bifurcation lines for $\Gamma^{(1)}=0.05, \Gamma^{(2)}=\Gamma^{(3)}=0$. (b) Saddle-node (solid lines) and Hopf bifurcation (dotted lines) in the limit of small damping $\propto \Gamma^{(1)}$. Not too close to the astroid (see Sec.~\ref{subsec:nonlocal_bifurcation}) for weak damping the system has the following states: (i) a stable and an unstable focus; (ii) two unstable foci and a stable limit cycle; (iii) a stable and an unstable focus and a stable and an unstable limit cycle; (iv) two stable foci and an unstable limit cycle.
}
\label{fig:astroid_gamma1}
\end{figure}

The saddle-node bifurcation curves for weak damping $\propto \Gamma^{(1)}$ are shown in Fig.~\ref{fig:astroid_gamma1}. Inside the curvilinear triangles the spin has four stationary states, whereas outside the triangles it has two stationary states.  In contrast to the case of damping $\propto \Gamma^{(3)}$ shown in Fig.~\ref{fig:astroid}, in the present case the bases of the triangles touch at $\mu=0$. From Eq.~(\ref{eq:trace_on_astroid}), one of the states emerging on the sides of the triangles is stable for $\mu > 0, |f|< 2^{-3/2}$ and is unstable otherwise; note that the stability changes in the middle of the bifurcation curves.

The occurrence of periodic oscillations of the spin is associated with Hopf bifurcations. In the present case, from Eq.~(\ref{eq:Hopf_general}) the Hopf bifurcational values of $s_z$ are $s_{zH}=\pm 1/\sqrt{2}$. Therefore Eq.~(\ref{eq:Hopf_through_s_H}) for the Hopf bifurcation lines for weak damping takes a simple form
\begin{eqnarray}
\label{eq:Hopf_explicit}
 f_H=2^{-1/2}\pm \mu, \quad f_H \notin (0,2^{-3/2});\\
 f_H=-2^{-1/2}\pm \mu, \quad f_H \notin (-2^{-3/2},0).\nonumber
\end{eqnarray}
These lines are shown in Fig.~\ref{fig:astroid_gamma1}(b). For $|f|\sim 1$ and far from the end points of the bifurcation lines, the typical frequency of the emerging oscillations is $\sim 1$  in dimensionless units, or $\sim DS/\hbar$ in dimensional units.

\subsection{Phase portrait far from the astroid}
\label{subsec:phase_portrait_gamma1}

Evolution of the spin phase portrait with varying parameters far away from the astroid, $|\mu|\gg 1$, can be understood by looking at what happens as the Hopf bifurcation curves are crossed, for example by varying $f$. The result is determined by two characteristics. One is stability of the stationary state for $f$ close to the bifurcational value $f_H$. The stability depends on the sign of ${\cal T}$ for small $f-f_H$ (note that ${\cal T}$ changes sign for $f=f_H$). The other characteristic is the sign of the quasienergy drift velocity $\langle \dot g\rangle$ for $f=f_H$ and for $g$ close to its bifurcational value $g_H$ at the stationary state. It is given by Eq.~(\ref{eq:energy_change}) [note that, generally, $\langle \dot g\rangle\propto (g-g_H)^2$ for $f=f_H$]. A combination of these characteristics tells on which side of the bifurcation point there emerges a limit cycle and whether this cycle is stable or unstable.

We write the value of $s_z$ at the Hopf bifurcation point as $ s_{zH}=\alpha/\sqrt{2}$, where $\alpha=\pm 1$, cf. Eq.~(\ref{eq:Hopf_general}). The bifurcational value of the field (\ref{eq:Hopf_explicit}) is $f_H=\pm\left(2^{-1/2}+\alpha \mu\right)\cos\phi_H$, where $\phi_H$ is the phase of the bifurcating stationary state.  Linearizing Eq.~(\ref{eq:trace}) in $s_z- s_{zH}$ and using the explicit form of the determinant ${\cal D}$ one can show that, for small $f-f_H$, in a stationary state $\sgn[{\cal T}/(f-f_H)]=-\sgn[\alpha f_H]$. Then 
\begin{eqnarray}
\label{eq:trace_Hopf}
  \sgn{\cal T}= -(\alpha\,\sgn f_H)\,\sgn(f-f_H).
\end{eqnarray}

The analysis of the quasienergy drift velocity near a Hopf bifurcation point is done in Appendix~\ref{sec:appendix_A}. It follows from Eqs.~(\ref{eq:ene_diss_Stocks_integral}), (\ref{eq:trace_average}) that
\begin{eqnarray}
\label{eq:energy_drift_near_Hopf}
 \langle \dot g\rangle = C\alpha\Gamma^{(1)}(g-g_H)^2\left(\beta |f_H|-\sqrt{2}\right),\nonumber\\
\sgn\left[\langle \dot g\rangle/(g-g_H)\right]=\alpha\beta\,\sgn\left(\beta |f_H|-\sqrt{2}\right),
\end{eqnarray}
where $C>0$ is a constant and $\beta=\sgn(f_H\cos\phi_H)\equiv \sgn(2^{-1/2}+\alpha\mu) = \pm 1$  [$\mu$ is related to $f_H$ by Eq.~(\ref{eq:Hopf_explicit})].

The sign of $\langle\dot g\rangle/(g-g_H)$ shows whether $g$ approaches $g_H$ as a result of damping or moves away from $g_H$. If $\sgn\left[\langle \dot g\rangle(g-g_H)\right] < 0$, the vicinity of the stationary state and the nascent limit cycle attracts phase trajectories. Therefore at a Hopf bifurcation a stable focus becomes unstable and a stable limit cycle emerges. On the other hand, if $\sgn\left[\langle \dot g\rangle(g-g_H)\right] > 0$, at a Hopf bifurcation an unstable focus transforms into a stable one and an unstable limit cycle emerges. 

Equation~(\ref{eq:trace_Hopf}) allows one to say on which side of $f_H$, i.e., for what sign of $f-f_H$ the stationary state is stable, since for a stable state ${\cal T}<0$. Therefore together Eqs.~(\ref{eq:trace_Hopf}) and (\ref{eq:energy_drift_near_Hopf}) fully determine what happens as $f$ crosses the bifurcational value.

We are now in a position to describe which states exist far from the astroid in different sectors (i)-(iv) in Fig.~\ref{fig:astroid_gamma1}(b). For small $|f|$ and large $|\mu|$, regions (i) in Fig.~\ref{fig:astroid_gamma1}(b), the system is close to a spin in thermal equilibrium, it has one stable and one unstable stationary state. We now start changing $f$ staying on the side of large positive $\mu$. When $f$ crosses one of the bifurcation curves $f_H=\pm \left(2^{-1/2}-\mu\right)$, the system goes to one of the regions (ii) in Fig.~\ref{fig:astroid_gamma1}(b). On the both bifurcation curves $\alpha=\beta =-1$. Therefore, from Eqs.~(\ref{eq:trace_Hopf}), (\ref{eq:energy_drift_near_Hopf}), when one of these curves is crossed as $|f|$ increases, there emerges a stable limit cycle, and the stable focus becomes unstable. As $|f|$ further increases it crosses the bifurcation curves $\pm(2^{-1/2}+\mu)$ and the system goes to one of the regions (iii) in Fig.~\ref{fig:astroid_gamma1}(b) (we assume that the crossing occurs in the region $|f_H|>2^{1/2}$). On these bifurcation curves $\alpha=\beta=1$. Therefore, from Eqs.~(\ref{eq:trace_Hopf}), (\ref{eq:energy_drift_near_Hopf}), when they are crossed with increasing $|f|$ there emerges an unstable limit cycle and the unstable focus becomes stable.

We now start from the range of large negative $\mu$ and small $|f|$. As we increase $|f|$ and cross the bifurcation curves $f_H=\pm(\mu + 2^{-1/2})$ the system goes from region (i) to one of the regions (iv) in Fig.~\ref{fig:astroid_gamma1}(b). From Eqs.~(\ref{eq:trace_Hopf}), (\ref{eq:energy_drift_near_Hopf}), in this case an unstable focus goes over into a stable focus and an unstable limit cycle emerges. Further crossing into one of the regions (iii) with increasing $|f|$ leads to a transformation of a stable focus into an unstable focus and an onset of a stable limit cycle. These arguments were used to establish the nomenclature of states in regions (i)-(iv) in Fig.~\ref{fig:astroid_gamma1}(b). They agree with the results of direct numerical calculations.

\subsection{Other bifurcations of limit cycles}
\label{subsec:nonlocal_bifurcation}

\subsubsection{Merging of stable and unstable limit cycles}
\label{subsubsec:limit_cycle_saddle}

\begin{figure}[h]
\includegraphics[width=2.4in]{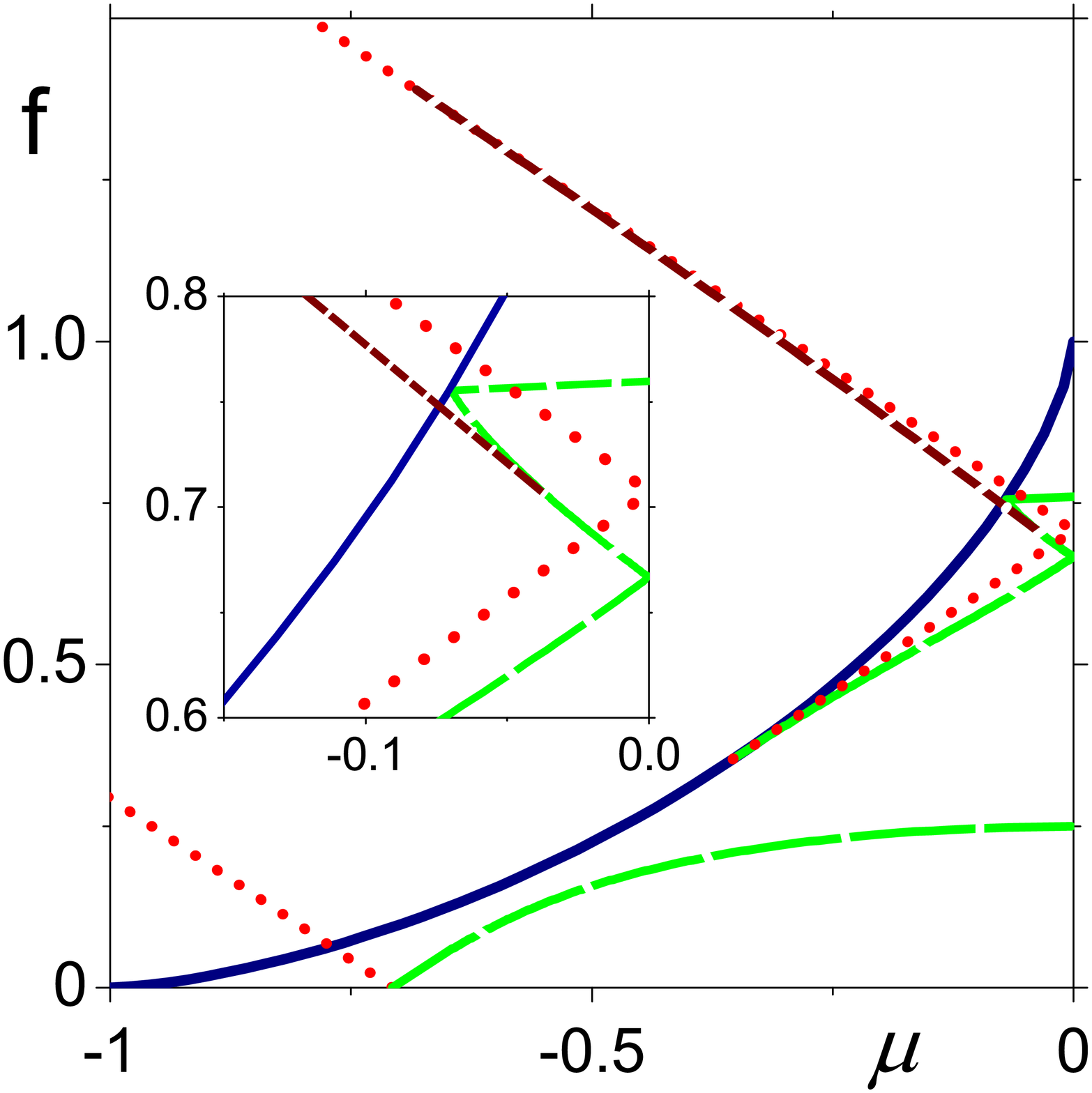}
\caption{(Color online). Bifurcation diagram in the limit $\Gamma^{(1)}\to 0$. The diagram is symmetric with respect to $\mu=0$ and $f=0$ axes, and therefore only the quadrant $f \geq 0, \mu \leq 0$ is shown. Saddle-node, Hopf, and saddle-loop bifurcation curves are shown by the solid, dotted, and long-dash lines, respectively, whereas the short-dash line shows the curve on which  stable and unstable limit cycles merge. }
\label{fig:full_bifurcation_gamma1}
\end{figure}

The number of periodic states in the rotating frame may change not only through Hopf bifurcations, but also through other bifurcations, where the radius of the bifurcating limit cycle does not go to zero. The simplest is a bifurcation where a stable limit cycle merges with an unstable limit cycle (saddle-node bifurcation of limit cycles). The onset of such bifurcations is clear already from Eq.~(\ref{eq:energy_drift_near_Hopf}). Indeed, at a Hopf bifurcation point the equation for the period-averaged quasienergy has a form $\langle \dot g\rangle = c(g-g_H)^2 + \ldots$ with $c\propto \beta|f_H|-\sqrt{2}$. For $|f_H|=\sqrt{2}$ on the bifurcation curves (\ref{eq:Hopf_explicit}) with $\beta = 1$ [the top and bottom dotted lines in Fig.~\ref{fig:astroid_gamma1}(b)] the coefficient $c=0$. This is a generalized Hopf bifurcation \cite{Guckenheimer1987}, see Fig.~\ref{fig:full_bifurcation_gamma1}.

At the generalized Hopf bifurcation, in phase space $(\phi, s_z)$ a stationary state merges simultaneously with a stable and an unstable limit cycle. In parameter space $(\mu,f)$, the Hopf bifurcation curve coalesces with the curve where stable and unstable limit cycles are merging, and the latter curve ends. The bifurcation curves are tangent, the distance between them scales as a square of the distance to the end point $\beta|f_H|=\sqrt{2}$ if the latter distance is small. This is seen in Fig.~\ref{fig:full_bifurcation_gamma1}. In the comparatively narrow region between the Hopf bifurcation curve and the corresponding limit-cycle merging curve the system has three limit cycles. One of these cycles disappears on the Hopf bifurcation curve, so that in regions (iii)  in Fig.~\ref{fig:astroid_gamma1}(b) there are two limit cycles and deeper in regions (ii) and (iv) there is one limit cycle. On its opposite end, the curve of merging limit cycles coalesces with the saddle-loop bifurcation curve.

\subsubsection{Saddle loops}
\label{subsubsec:saddle_loops}

Spin dynamics for damping $\propto \Gamma^{(1)}$ is characterized also by global bifurcations of the type of saddle loops. This is clear already from the analysis of the end points of the Hopf bifurcation curves. These points lie on the curves of saddle-node bifurcations. The corresponding equilibrium point has double-zero eigenvalue, and the behavior of the system near this point is well-known \cite{Guckenheimer1987}. The Hopf bifurcation curve is tangent to the saddle-node bifurcation curve at the end point. In addition, there is a saddle-loop bifurcation curve coming out of the same end point and also tangent to the saddle-node bifurcation curve at this point. At a saddle-loop bifurcation the system has a homoclinic trajectory that starts and ends at the saddle point.

\begin{figure}[h]
\includegraphics[width=2.4in]{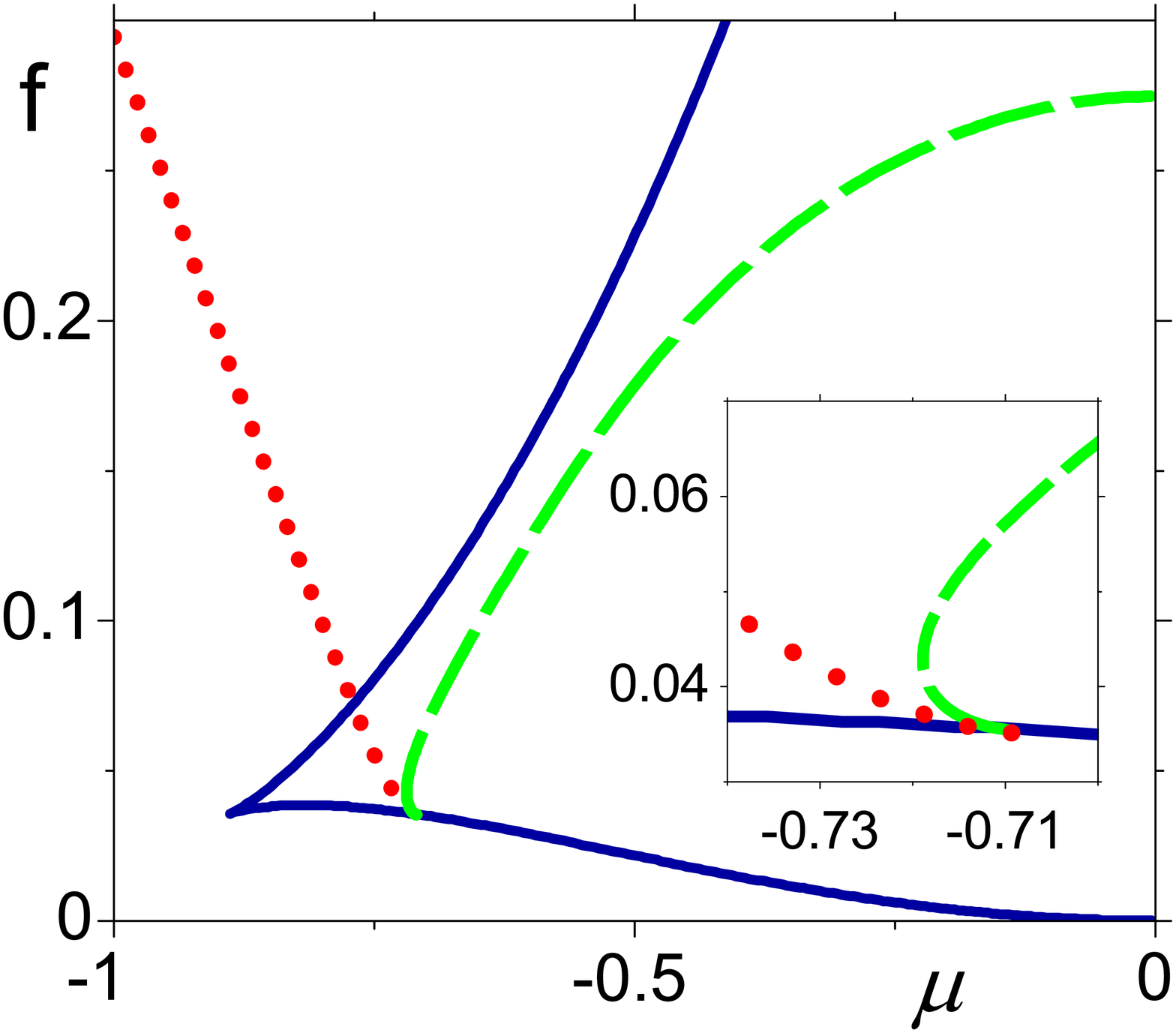}
\caption{(Color online). Bifurcation diagram near the end point of the Hopf bifurcation line which in the limit $\Gamma^{(1)}\to 0$ has the form $f_H=-\mu-2^{-1/2}$. For nonzero $\Gamma^{(1)}$ this bifurcation line ends on the saddle-node bifurcation line (\ref{eq:astroid_bases}). The plot refers to  $\Gamma^{(1)} = 0.0125$. The inset shows a close vicinity of the end point. Hopf, saddle-node, and saddle loop bifurcation curves are shown by dotted, solid, and long-dashed lines, respectively. Other Hopf bifurcation curves that go to $f_H=0$ for $\Gamma^{(1)}\to 0$ display a similar behavior near their end points. }
\label{fig:detailed_Hopf_small_f}
\end{figure}

The structure of vicinities of the end points of the Hopf bifurcation curves is shown in Figs.~\ref{fig:full_bifurcation_gamma1} and \ref{fig:detailed_Hopf_small_f} for the curves ending on the sides and the bases of the saddle-node bifurcation triangles, respectively. Note that the Hopf bifurcation curves that crossed at $f=0$ in the limit $\Gamma^{(1)}\to 0$ are separated for finite $\Gamma^{(1)}$. They end on the saddle-node bifurcation curves.

We have found numerically a fairly complicated pattern of saddle-loop bifurcation curves. Full analysis of this pattern is beyond the scope of this paper.

\subsection{Hysteresis of spin response in the presence of limit cycles}
\label{subsec: hysteresis_gamma1}

\begin{figure}[h]
\includegraphics[width=3.2in]{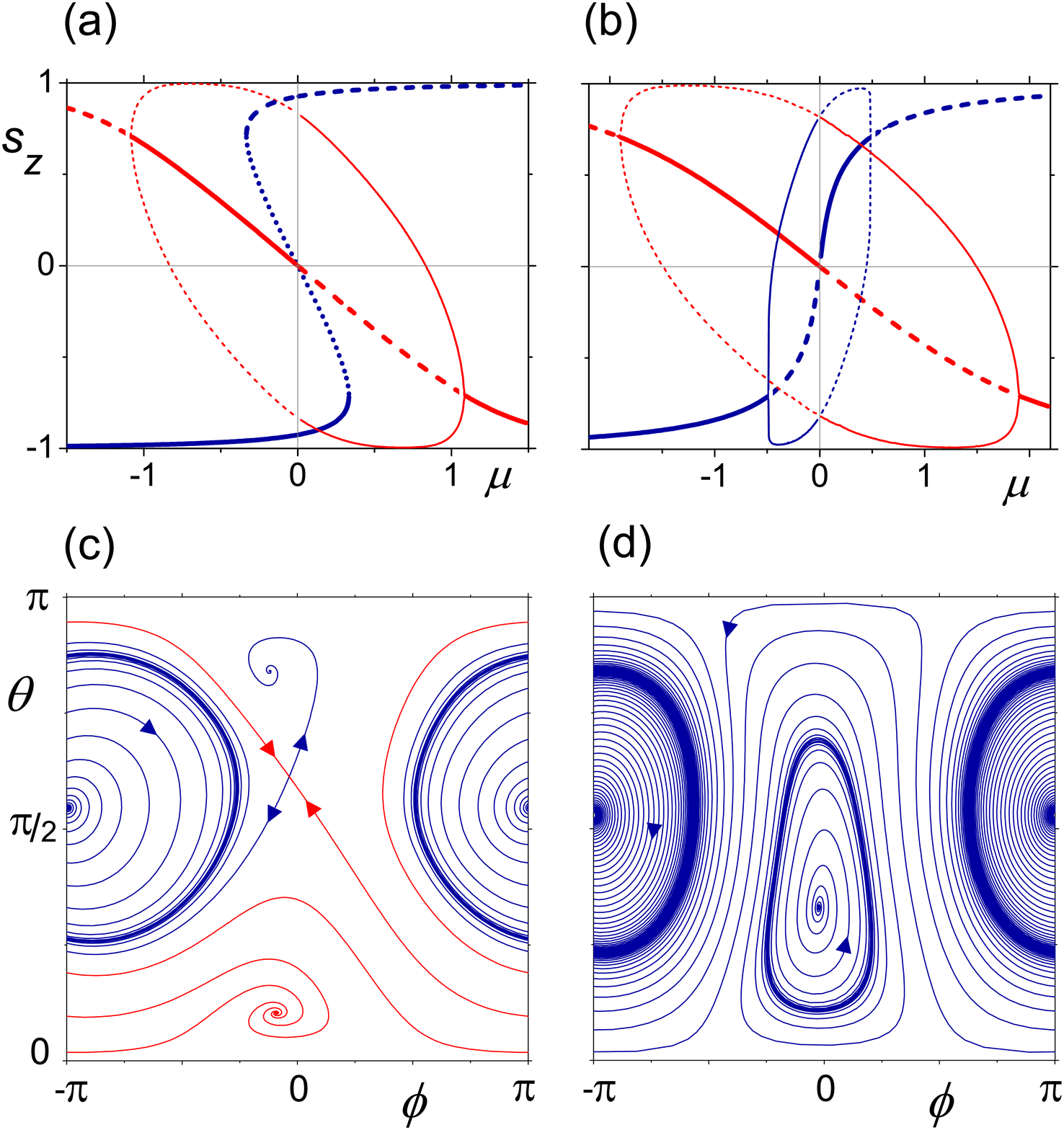}
\caption{(Color online). Panels (a) and (b): hysteresis of the spin dynamics with varying scaled detuning of the modulating field frequency $\mu$. In (a) $f=0.4$, so that $\mu$ goes through the curvilinear bifurcation triangle in Fig.~\ref{fig:astroid_gamma1}. In (b) $f=1.2$, it lies above the triangles. Bold solid lines, long dashed lines, and dotted line show stable and unstable equilibria and the saddle stationary state, respectively. Pairs of thin solid lines and short dashed lines show the boundaries (with respect to $s_z$) of stable and unstable limit cycles. Panels (c) and (d): phase portraits for $\mu=0.2$. In (c) and (d) $f=0.4$ and $1.2$, respectively. The arrows show the direction of motion along the trajectories. The data refer to $\Gamma^{(1)}=0.05$. }
\label{fig:hyster_gamma1}
\end{figure}

Coexistence of stable stationary states and stable limit cycles in the rotating frame leads to hysteresis of the response of a spin when the modulating field parameters are slowly varied. Examples of such hysteresis with varying scaled frequency detuning $\mu$ and the characteristic phase portraits are shown in Fig~\ref{fig:hyster_gamma1}.

The hysteretic behavior is unusual. This is a consequence of the feature of the spin dynamics for $\mu=0$ where either all phase trajectories are closed loops (for $f$ outside the curvilinear saddle-node bifurcation triangles in Fig.~\ref{fig:astroid_gamma1}) or all trajectories in a part of the phase plane are closed loops (for $f$ inside the triangles in Fig.~\ref{fig:astroid_gamma1}). As a result two or more states (stationary or periodic) simultaneously loose or acquire stability as $\mu$ goes through $0$. This leads to an ambiguity of switching, a ``Buridan's ass" type situation. Where a stable branch looses stability for $\mu=0$, the system has more than one stable state to switch to. Also, in contrast to the situation of Sec.~\ref{sec:linear_in_S} where the system had no limit cycles, hysteresis emerges whether the varying field parameter does or does not cross the saddle-node bifurcation lines.

Figures~\ref{fig:hyster_gamma1}(a) and (b) show the behavior of the system with varying $\mu$ for $f$ inside and outside the saddle-node bifurcation triangles, respectively. It should be noted that we chose $f$ in Fig.~\ref{fig:hyster_gamma1}(a) so that the saddle-loop bifurcation line is not encountered, which provides an insight into the most basic features of the hysteresis. In addition, in Fig.~\ref{fig:hyster_gamma1}(b) we do not show an extremely narrow region near Hopf bifurcation lines $\mu\approx \pm (f-2^{-1/2})$ where the system has small-radii stable and unstable cycles which merge on the short-dash bifurcation line in Fig.~\ref{fig:full_bifurcation_gamma1}.

In Fig.~\ref{fig:hyster_gamma1}(a), for large negative $\mu$ the system has one stable state (with negative $s_z$). As $\mu$ increases this state disappears via a saddle-node bifurcation and the system switches to a stable limit cycle. For chosen $f=0.4$ this happens for $\mu \approx 0.33$. With further increase of $\mu$ the limit cycle shrinks and ultimately disappears via a Hopf bifurcation, and then the stationary state inside the cycle becomes stable. 

On the other hand, if we start in Fig.~\ref{fig:hyster_gamma1}(a) from large positive $\mu$ and decrease $\mu$, the stable stationary state via a supercritical Hopf bifurcation becomes a stable limit cycle. The cycle looses stability at $\mu=0$, and as $\mu$ becomes negative the system can switch either to the stable stationary state inside the cycle (with $s_z \to +0$ for $\mu\to -0$) or to a stable stationary state outside the cycle with negative $s_z$. The stable state with $s_z\to +0$ for $\mu\to -0$, ultimately looses stability with decreasing $\mu$  via a Hopf bifurcation (at $\mu \approx -f-2^{-1/2}$, for small damping, cf. Fig.~\ref{fig:astroid_gamma1}). If the system is in this state, it switches to the stable equilibrium with negative $s_z$.

A typical phase portrait for $f=0.4, 0<\mu<0.33$ is shown in Fig.~\ref{fig:hyster_gamma1}(c). It gives an insight into the behavior described above. The system has a stable limit cycle with an unstable focus inside and with stable and unstable equilibria and a saddle point outside the limit cycle. For $\mu = 0$ the system has a homoclinic saddle connection, and all trajectories inside the homoclinic trajectory are closed loops, cf. Fig.~\ref{fig:small_gamma_2_traj}(c)

In Fig.~\ref{fig:hyster_gamma1}(b), for large negative $\mu$ the system also has one stable state (with negative $s_z$). As $\mu$ increases this state looses stability via a Hopf bifurcation (at $\mu \approx -f + 2^{-1/2}$, for small damping). The emerging state of stable oscillations looses stability for $\mu=0$. For larger $\mu$ the system switches either to the stationary state inside the limit cycle (with $s_z\to +0$ for $\mu\to +0$) or to another stable periodic state. The coexistence of stable and unstable limit cycles with stationary states inside of them is seen in Fig.~\ref{fig:hyster_gamma1}(d). 

As $\mu$ becomes positive and further increases, the stable stationary state inside the unstable cycle looses stability by merging with this cycle, and the system switches to the periodic state corresponding to the stable limit cycle in Fig.~\ref{fig:hyster_gamma1}(d). For still larger $\mu$ ($\mu\approx f+2^{-1/2}$, for weak damping) this state experiences a Hopf bifurcation and becomes a stable stationary state. The behavior with $\mu$ decreasing from large positive values can be understood from Fig.~\ref{fig:hyster_gamma1} in a similar way.

\section{Conclusions}
\label{sec:conclusions}

We have developed a microscopic theory of a resonantly modulated large spin in a strong static magnetic field and studied spin dynamics in the classical limit. We have taken into account relaxation processes important for large-spin systems of current interest. They correspond to transitions between neighboring and next-neighboring Zeeman levels with emission or absorption of excitations of a bosonic thermal bath. Classical spin dynamics depends significantly on the interrelation between the rates of different relaxation processes. Generally it is not described by the Landau-Lifshitz equation for magnetization in a ferromagnet, although one of the coupling mechanisms that we discuss leads to the Landau-Lifshitz damping in the rotating frame.

We found that the spin dynamics has special symmetry at exact resonance where the modulation frequency is equal to the Larmor frequency, $\omega_F=\omega_0$. This symmetry leads to a Hamiltonian-like behavior even in the presence of dissipation. In the rotating frame, phase trajectories of the spin form closed loops in a part of or on the whole phase plane. Therefore when $\omega_F$ goes through $\omega_0$ several states change stability at a time.

The simultaneous stability change leads to unusual observable features. Where the system has only one stable state for a given parameter value, as $\omega_F$ goes through $\omega_0$ there occurs switching between different states that leads to an abrupt change of the magnetization. The behavior is even more complicated where several stable states coexist for $\omega_F$ close but not equal to $\omega_0$. Here, where $\omega_F-\omega_0$ changes sign, the state into which the system will switch is essentially determined by fluctuations or by history (if $\omega_F$ is changed comparatively fast).

We found the conditions where the spin has more than one stable stationary state in the rotating frame. Such stable states correspond to oscillations of the transverse magnetization at the driving frequency in the laboratory frame. Multistability leads to magnetization hysteresis with varying parameters of the modulating field. If the fastest relaxation process is transitions between neighboring states due to coupling quadratic in spin operators, the resonantly modulated spin can have periodic nonsinusoidal states in the rotating frame with frequency $\propto DS/\hbar$, where $D$ is the anisotropy energy. In the laboratory frame, they correspond to oscillations of the transverse magnetization at combinations of this frequency (and its overtones) and the Larmor frequency.

Quantum fluctuations of the spin lead to phase diffusion of the classical periodic states in the rotating frame. As a result, classical oscillations decay. The intensity of quantum fluctuations and the related decay rate depend on the value of $S^{-1}$. We have found \cite{Hicke2008} that the oscillations decay comparatively fast even for $S=10$. Therefore they are transient. Still the classically stable vibrations lead to pronounced features of the full quantum spin dynamics.

The present analysis can be immediately extended to a more general form of the spin anisotropy energy, in particular to the case where along with $DS_z^2$ this energy has a term $E(S_x^2-S_y^2)$, which is important for some types of single-molecule magnets \cite{Gatteschi2006}. In the RWA, the corresponding term renormalizes $D$ and $\omega_0$. The analysis applies also to decay due to two-phonon or two-magnon coupling, which often plays an important role in spin dynamics and leads to energy relaxation via inelastic scattering of bath excitations by the spin. Another important generalization is that the results are not limited to linearly polarized radiation. It is easy to show that they apply for an arbitrary polarization as long as the radiation is close to resonance.

In conclusion, starting from a microscopic model, we have shown that the classical dynamics of a resonantly modulated large spin in a strong magnetic field displays several characteristic features. They include abrupt switching between magnetization branches with varying parameters of the modulating field even where there is no hysteresis, as well as the occurrence of hysteresis and an unusual pattern of hysteretic inter-branch switching. These features are related to the Hamiltonian-like behavior of the dissipative spin for modulation frequency equal to the Larmor frequency in the neglect of spin anisotropy. Along with forced vibrations at the modulation frequency, the transverse spin components can display transient vibrations at a combination of the modulation frequency and a slower frequency $\propto DS/\hbar$ and its overtones.  They emerge if the fastest relaxation mechanism corresponds to transitions between neighboring Zeeman levels with the energy of coupling to a thermal bath quadratic in the spin operators.

We are grateful to S.~W.~Shaw for the discussion of the bifurcation pattern and to J.~Vidal for pointing to the analogy with the Lipkin-Meshkov-Glick model. This research was supported in part by the NSF through grant PHY-0555346 and by the Institute for Quantum Sciences at MSU.


\appendix

\section{Energy change near a Hopf bifurcation}
\label{sec:appendix_A}

In this Appendix we outline the calculation of the relaxation of quasienergy $g$ near a Hopf bifurcation point. For concreteness we assume that $\Gamma^{(2)}= \Gamma^{(3)} = 0$ and the only nonzero damping parameter is $\Gamma^{(1)}$. For small damping a stationary state that experiences a bifurcation has phase $\phi_H$ close to either 0 or $\pi$, whereas $ s_{zH}\approx \pm 2^{-1/2}$. The dynamics is characterized by two parameters, $\alpha=\sgn s_{zH}$ and $\beta=\sgn[f_H\cos\phi_H]$. The bifurcational value of the field for $\Gamma^{(1)}\to 0$ is $f_H= (2^{-1/2}+\alpha\mu)\cos\phi_H$ [cf. Eq.~(\ref{eq:Hopf_explicit})].

At the bifurcating stationary state the quasienergy is $g_H=g\left(\phi_H, s_{zH}\right)$; it is easy to see that this is a local minimum of $g(\phi,s_z)$ for $\beta >0$ or a maximum for $\beta < 0$. Respectively, on phase plane $(\phi, s_z)$ the constant-$g$ trajectories close to the bifurcating stationary state rotate about this state clockwise for $\beta > 0$ and counterclockwise for $\beta < 0$. The angular frequency of this rotation is $\approx 2\pi/\tau_p(g_H) = {\cal D}^{1/2}$, where ${\cal D}$ is given by Eq.~(\ref{eq:determinant}).

We now consider dissipation-induced drift over quasienergy $\langle \dot g\rangle$. It is given by Eq.~(\ref{eq:energy_change}). Noting that  $\partial _{s_z}g=\dot \phi$ and using the Stokes theorem we can rewrite this equation as
\begin{eqnarray}
\label{eq:ene_diss_Stocks_integral}
 \langle \dot g\rangle=\beta\tau_p^{-1}(g)\int\,d\phi\,ds_z {\cal T},
 \end{eqnarray}
where the integral is taken over the interior of the constant-$g$ orbit on the $(\phi,s_z)$ plane and ${\cal T}\equiv {\cal T}(s_z)$ is given by Eq.~(\ref{eq:trace}). At a Hopf bifurcation point ${\cal T}=0$. Therefore ${\cal T}(s_z)$ in Eq.~(\ref{eq:ene_diss_Stocks_integral}) must be expanded in $\delta s_z=s_z- s_{zH}$.

It is convenient to calculate integral (\ref{eq:ene_diss_Stocks_integral}) by changing to integration over action-angle variables $(I,\psi)$, which are canonically conjugate to $(s_z,\phi)$, with $g$ being the effective Hamiltonian. The angle $\psi$ gives the phase of oscillations with given quasienergy $g$. The action variable $I=(2\pi)^{-1}\oint s_zd\phi$ is related to $g$ by the standard expression $dI/dg= \tau_p(g)/2\pi\approx {\cal D}^{-1/2}$; we note that $I$ becomes negative away from the stationary state for $\beta < 0$.

In evaluating expression (\ref{eq:ene_diss_Stocks_integral}) it is further convenient to start with integration over $\psi$. It goes from 0 to $2\pi$ and corresponds to period averaging for a given $I\propto \delta g= g-g_H$ (integration over $I$ corresponds to integration over $\delta g$).

If vibrations about $\bigl(\phi_H, s_{zH}\bigr)$ were harmonic, the lowest-order term in $\delta s_z$  that would not average to zero on integration over $\psi$ would be $(d^2{\cal T}/d s_z^2) (\delta s_z)^2/2\propto |\delta g|$ (the derivative of ${\cal T}$ is calculated at the bifurcating stationary state). However, it is easy to see that the integral over $\psi$ of the linear in $\delta s_z$ term in ${\cal T}$ is also $\sim\delta g$. It can be calculated from equation of motion $\dot\phi=\partial_{s_z}g$ by expanding the right-hand side to second order in $\delta s_z,\delta\phi$ and noting that $\overline{\dot\phi}=0$, where the overline means averaging over $\psi$. This gives, after some algebra,
\begin{eqnarray}
\label{eq:trace_average}
 \bar{\cal T} =&& 64\Gamma^{(1)}\alpha\,(\delta g)\left(2^{3/2}\beta |f_H|-1\right)^{-2}\nonumber\\
&&\times\left(\beta|f_H|-2^{1/2}\right).
\end{eqnarray}
This expression combined with Eq.~(\ref{eq:ene_diss_Stocks_integral}) shows how the energy relaxation rate depends on the field $f_H$. It is used in Section~\ref{sec:quadratic_in_S} to establish the full bifurcation diagram.



\end{document}